# Fractional Order Phase Shaper Design with Routh's Criterion for Iso-damped Control System

Suman Saha[1], Saptarshi Das[1], Ratna Ghosh[2], Bhaswati Goswami[2], Amitava Gupta[1]
1. Power Engineering Department, Jadavpur University,
2. Instrumentation & Electronics Department, Jadavpur University, Kolkata-700098, India.
email: the_suman@yahoo.co.in

R. Balasubramanian[3], A.K. Chandra[3], Shantanu Das[4]
3. R & D, Electronic System, Nuclear Power Corporation of India Ltd.,
4. Bhabha Atomic Research Centre, Mumbai-400085, India.

*Abstract*—**Phase curve of an open loop system is flat in nature if the derivative of phase with respect to frequency is zero. With a flat phase curve, the corresponding closed-loop system exhibits an iso-damped property i.e. maintains constant overshoot with the change of gain and with other parametric variations. In recent past application, fractional order (FO) phase shapers have been proposed by contemporary researchers to achieve enhanced parametric robustness. In this paper, a simple Routh tabulation based methodology is proposed to design an appropriate FO phase shaper to achieve phase flattening in a control loop, comprising a system, controlled by a classical PID controller. The method is demonstrated using MATLAB simulation of a second order DC motor plant and also a first order with time delay system.**

## I. INTRODUCTION

TO make a system robust from gain (parametric spreads) variation and to keep phase margin constant, it is necessary to have phase derivative with respect to frequency to be zero around the gain crossover frequency. However, a flat phase tuning method like the one presented in [1] cannot make the phase curve flat about a wide range of frequency and also cannot determine the width of the flat phase region. The contribution of this paper is to use Carlson's approximation [2], [3] of fractional integrators and differentiators (differ-integrators) as phase shaping elements to obtain iso-damped step response of process plants satisfying flat phase condition. A new concept of designing a FO phase shaper using Routh's stability criterion [4], [5] with iso-damped step response of a first and second order plant with and without time delay, controlled by a PID controller, simulated by MATLAB/Simulink is presented. With the plant $G_p(s)$ and controller $G_c(s)$ known, a FO phase shaper $G_{ph}(s)$ is designed to achieve flattening of the asymptotic phase curve about a wide range of frequency, which ensures that the closed loop system will exhibit an iso-damped step response. The phase shaper is designed for a plant model with a classical PID controller tuned by some standard techniques, like LQR based one, as proposed by He *et al* [6]. A comprehensive survey of different PID tuning methods found in [7], [8], [9]. The approximation method for the FO phase shaper is based on Carlson's approximation and modified for a common structure.

The methodology to get the first order compensator [10] model as a phase shaper is first tried with a practical system represented by a DC motor [11], controlled by a PI controller and the first order plus time delay (FOPTD) [12] system controlled by a PID controller. The Fractional order phase shaper transfer function from the compensator model can be estimated but the use of a first order phase shaper approximation rather than the higher order one [3], representing an exact fractional order, along with a PID controller for a given plant, is extremely advantageous from practical considerations as such a hardware can be readily used with the existing PID controller associated with a plant.

The paper is organized as follows. Section II presents an overview of Carlson's method for representation of a FO differentiator & integrator and the concept of phase shaper. Routh-stability criterion based design of a FO phase shaper is presented next, along with simulation examples in Section III, followed by conclusion in Section IV.

## II. INTEGER REPRESENTATION OF FRACTIONAL ORDER ELEMENTS

### A. *Carlson's approach*

The Carlson's approach using Newton's approximation [2], [3] to find arbitrary $q^{\text{th}}$, $(0 \le q \le 1)$ root of polynomial is first described. Thus if $H(s)$ be the $q^{\text{th}}$ root of a polynomial $G(s)$, then $H(s) = (G(s))^q$ (1)

where $$q = \frac{m}{p} \quad (2)$$

$p$ and $m$ are integers. Defining, $H_0(s) = 1$, the value of $H(s)$ at the $i^{\text{th}}$ iteration can be represented by

$$H_i(s) = H_{i-1}(s)\frac{(p-m)(H_{i-1}(s))^2 + (p+m)G(s)}{(p+m)(H_{i-1}(s))^2 + (p-m)G(s)} \quad (3)$$

Thus, the first order Carlson's approximation of a simple FO integrator can be represented as

$$s^{-q} = \frac{s + \frac{(p+m)}{(p-m)}}{\frac{(p+m)}{(p-m)}s + 1} \quad (4)$$

The ratio of $\frac{(p+m)}{(p-m)} = \alpha$ is taken and the equation can be rewritten as $s^{-q} = \frac{s+\alpha}{\alpha s + 1}, \alpha > 1$ (5)

where the order of the integrator is defined by

$$q = \frac{(\alpha - 1)}{(\alpha + 1)} \quad (6)$$

Following the same treatment as presented above, a FO differentiator can be represented as

$$s^{q} = \frac{s+\alpha}{\alpha s + 1}, \alpha < 1 \quad (7)$$

The order of the differentiator is

$$q = \frac{(1-\alpha)}{(1+\alpha)} \quad (8)$$

Therefore, $\frac{s+\alpha}{\alpha s+1}$ , from equation (5, 7) is the common first order approximate transfer function of a FO differ-integrator, and the knowledge of $\alpha$ ( $\alpha$<1 or $\alpha$>1) enables to transfer a fractional differentiator or integrator from the above transfer function using equation (6, 8).

### *B. Phase shaping with FO differ-integrator*

It is seen that a FO differ-integrator $s^q$ adds a phase of $\frac{\pi q}{2}$ to the asymptotic phase curve. When $q$ is positive, this results in a phase boost and when $q$ is negative, this causes the phase to buck or drooping in nature. Thus, a differ-integrator can be used for phase addition or subtraction to modify the transient response of a closed loop system with a given controller in loop [2].

However, from practical considerations, a phase shaper has to be represented by a rational transfer function of small order to be practically realizable and hence the actual phase contribution of $s^q$ is perceptible over a finite frequency spread. It is clearly seen that the phase contribution by $s^q$ realized by the first order Carlson's approximation is limited to the frequency interval $\left[\alpha, \frac{1}{\alpha}\right]$ for a FO differentiator and $\left[\frac{1}{\alpha}, \alpha\right]$ for a FO integrator. In fact, it is seen that a phase shaper represented by first order Carlson's representation acts like a compensator.

### *C. Concept of phase shaping*

The methodology proposed by Chen *et al.* in [1] ensures increased parametric robustness by altering the shape of the phase curve at the tangent frequency. This basic robust PID tuning methodology has been modified by Chen *et al.* in [1] which has been used with the PID controller tuned by the methodology reported in [6], [7], [8]. The lower cut-off frequency of the flat-phase region is defined by the PID controller and the upper cut-off frequency by the phase shaper. A phase shaper should have the following characteristics [1]:

(1) It should be of lower order polynomial in both numerator and denominator for approximation of a particular fractional order.

(2) The resultant phase using the phase shaper should be constant about a wide range of frequency including gain-crossover frequency.

It may be mentioned that introduction of a FO differ-integrator $s^q = \frac{s+\alpha}{\alpha s+1}$ modifies the phase curve around nominal frequency $\omega_0 = 1$ rad/s. Now, in order to shift the nominal frequency from $\omega_0$ to any arbitrary frequency $\omega_r$ , where extra phase has to be added or reduced, it is necessary to consider the altered FO phase-shaper of the form $K(s^q + a)$, which can be realized by a first order Carlson's approximation about the frequency range $\left[\frac{1}{\alpha}, \frac{\alpha + a}{1+\alpha a}\right]$ as

$$K(s^q + a) = K\frac{(1+a\alpha)s + (a+\alpha)}{\alpha s + 1} \quad (9)$$

It modifies the phase curve such that the peak occurs at the frequency $\omega_r = \sqrt{\frac{\alpha + a}{\alpha(1+\alpha a)}}$ (10)

The modified phase shaping transfer function can also be of the form $K(s+a)^q = K\left(\frac{s+\alpha+a}{\alpha s + 1 + \alpha a}\right)$ (11)

which modifies the phase by $\phi_m = \tan^{-1}\left(\frac{1-\alpha^2}{2\alpha\omega_r}\right)$ (12)

about $\omega_r = \sqrt{\frac{(\alpha+a)(1+\alpha a)}{\alpha}}$ where $\alpha, a > 0$ (13)

### *D. Routh array based design*

In this section Routh criterion based methodology is proposed for phase shaping to modify the resultant-phase curve to be flattened about gain cross-over point. The Routh–Hurwitz criterion states that the number of roots of a polynomial with positive real parts is equal to the number of changes of sign in the first column of the array. This criterion requires that there be no sign change in the first column for a stable system. This requirement is both necessary and sufficient [4]. Not only Routh array gives stability analysis, there also exists a direct relationship between the stability criterion and gain margin of the system in terms of stability ratio (SR) [5].

$$SR = \frac{K_m}{K}, \quad (14)$$

$K_m$ is the marginal gain & $K$ is the actual gain.

$K_m$ can be obtained from the $s^1$ row test function of the Routh array. Phase margin can be obtained from $s^2$ row element and the gain margin also from the relation

$$\omega_{gc} = \omega_{pc}\left[\frac{K}{K_m}\right]^{0.5} \quad (15)$$

In our proposed method a simple phase shaper of the form $s^q = \frac{s+\alpha}{\alpha s+1}$ is introduced in the loop so that the characteristic equation of the resultant closed-loop system can be expressed as

$$1 + K\left(\frac{1}{\alpha}\right).\left(\frac{s+\alpha}{\alpha s+1}\right).\left(K_p + \frac{K_i}{s} + K_d s\right).G_p(s) = 0 \quad (16)$$

Now searching the best value of $\alpha$ that maximizes $K$ from the $s^1$ row test function of the routh array or get highest stability margin $K_m$ to produce a stable closed-loop system. By following the argument presented, this flattens the phase-curve, thus producing a system with constant overshoot (iso-damped response) over a considerable gain-spread, as the phase margin remains fairly constant.

## III. Simulation and Results

### A. Control of DC motor

An example of speed control of a DC motor [11] is taken, with the following chosen parameters:

Moment of inertia of the rotor $J = 0.01$ kg.m²/s²,
Damping ratio of the mechanical system $b = 0.1$ Nms,
Electromotive force constant $K_e = K_t = 0.01$ Nm/Amp,
Electric resistance $R = 1$ ohm,
Electric inductance $L = 0.5$ H

For that the transfer function of the plant is

$$G_p(s) = \frac{0.01}{0.005s^2 + 0.06s + 0.1001} \quad (17)$$

It is seen that the plant is an overdamped system and a LQR based tuning methodology introduced by He *et al.* [6] to produce a closed loop damping $\zeta_{cl} = 0.8$ and $\omega_{cl} = 3.0 rad/s$, yields $K_P = 1.64$, $K_I = 2.64$, $K_D = 0$

Thus, the transfer function of a PI controller obtained is

$$G_c(s) = 1.64 + \frac{2.64}{s} \quad (18)$$

Thereafter, a phase shaper with a transfer function of $\frac{s+\alpha}{\alpha s+1}$, from equation (5,7) is taken as a realizable approximation of a differ-integrator and the corresponding Routh array becomes

| | | | |
|---|---|---|---|
| $s^4$ | .005α | (.1001α+.06+.0164K) | .0264αK |
| $s^3$ | .06α+.005 | (.1001+.0264K+.0164αK) | 0 |
| $s^2$ | A | B | 0 |
| $s^1$ | C | 0 | |
| $s^0$ | B | | |

Thus the objective is to find out the value of $\alpha$ or the order of a phase shaper ($q$) which maximizes the loop gain or $K_m$, satisfying the following constrains:

$$A, B, C > 0 \quad (19)$$

where $A, B$ and $C$ comes from Routh array.

Satisfying all the constrains of (19), $\alpha = 0.5$ corresponds to the maximum value of $K_m$. Hence this value of $\alpha$ is chosen and the phase shaper becomes

$$G_{ph1}(s) = \frac{s+0.5}{0.5s+1} \quad (20)$$

This is a FO differentiator with the order of $s^{0.33}$ using equation (7, 8). As mentioned earlier, the first order realization of phase shaper $s^{0.33}$ has not made the flat phase around the crossover point. So the extra phase is compensated by two other FO integrators using equations (12, 13) and the net transfer function becomes

$$G_{ph}(s) = G_{ph1}(s) \times G_{ph2}(s) \times G_{ph3}(s) = \frac{s^{0.33}}{(s+5.8)^{0.8}(s+58)^{0.99}} \quad (21)$$

where, $G_{ph2}(s) = \frac{1}{(s+5.8)^{0.8}}, G_{ph3}(s) = \frac{1}{(s+58)^{0.99}}$

The resultant phase curve is shown in fig. 1 with thin line.

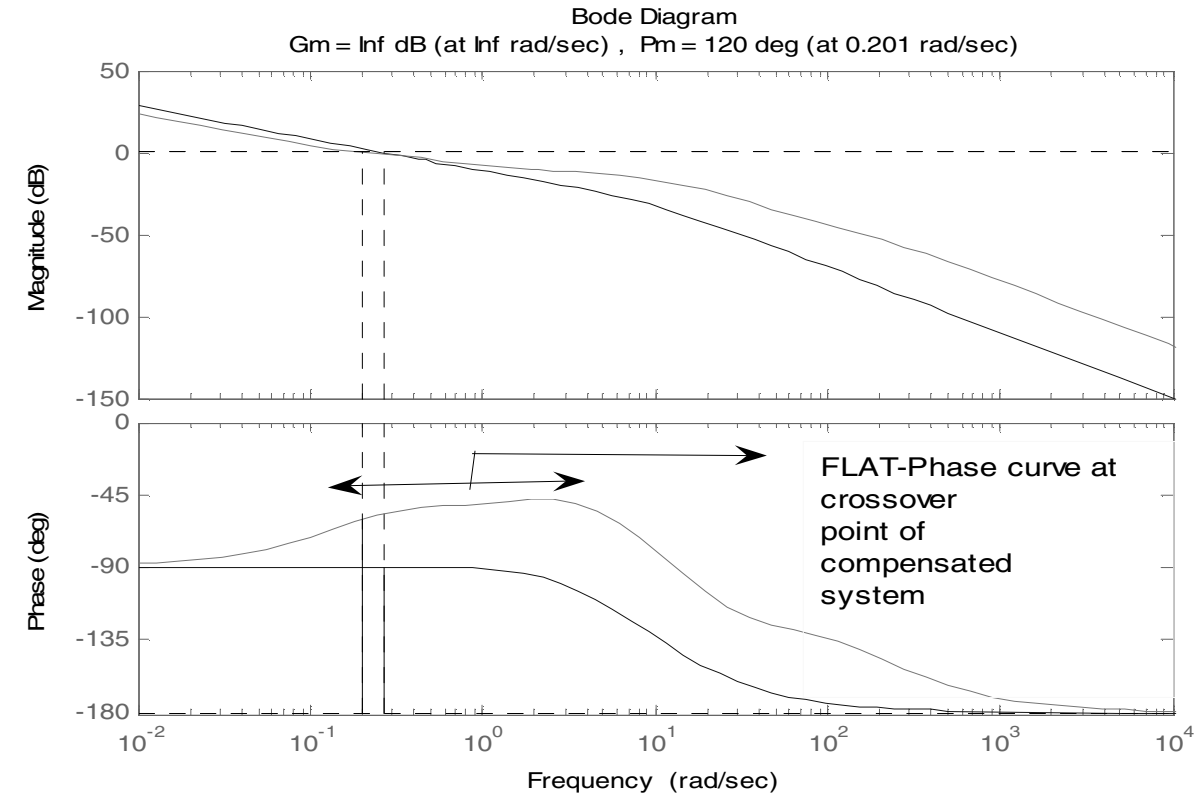

Fig. 1. Magnitude and phase plot with the phase adjustments

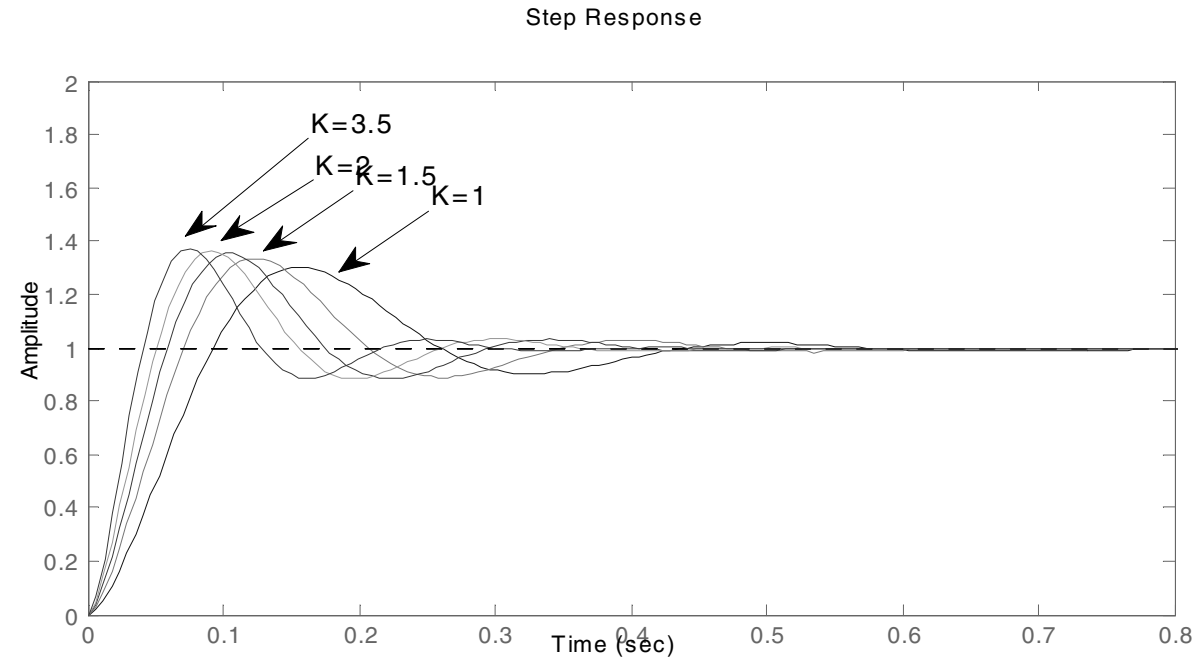

Fig. 2. Iso-damped step response with the variation of loop gain

## B. *Control of time delay system*

$$G(s) = \frac{5}{1.5s+1} e^{-s} \qquad (22)$$

An balanced type of plant ( $L \simeq T$ ) [12] is controlled by a PID controller [6]:

$$G_c(s) = 0.364 + \frac{0.22}{s} + 0.149s \qquad (23)$$

The frequency response of the corresponding open loop plant with controller is given in fig. 3. A first order Carlson's approximation of equation (9) used to get the order of the shaper as

$$G_{ph}(s) = \frac{1}{s^{0.66}+1} \qquad (24)$$

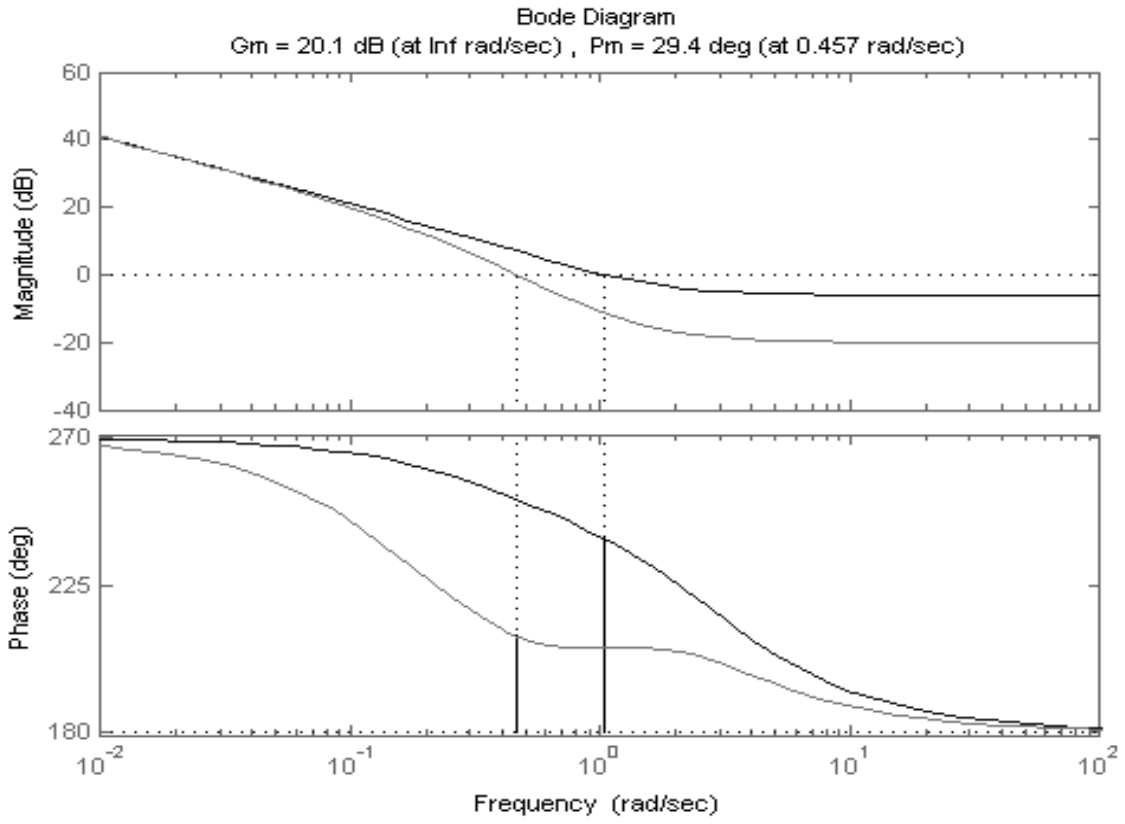

Fig. 3. Flat phase using First order phase shaper transfer function

This phase shaper flattens the phase curve around $\omega_{gc}$ (gain crossover frequency) of the plant model along with the PID controller. Fig. 4 shows the corresponding time response of the plant & PID controller along with a higher order realized phase shaper subjected to a unit step input to check whether the fractional order phase shaper also exhibits iso-damped behavior or not. The result shows constant overshoot for considerable amount of gain variation (20%).

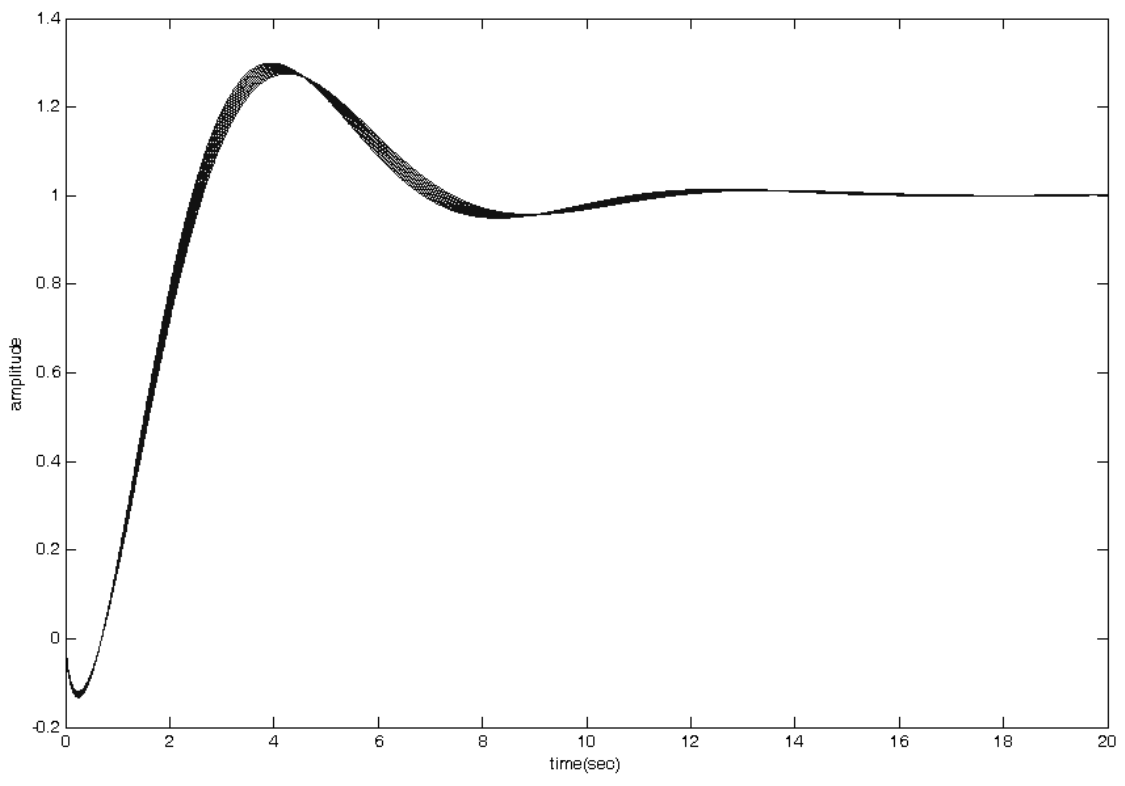

Fig. 4. Step responses with 20% gain variation using higher order realized model of phase shaper using Carlson's realization method.

## IV. Conclusion

The methodology put forward in this paper, can be used for enhancing the robustness of PID control loops. It is well known that PID controller is predominantly used for process control applications. More than 90% of the controllers are PID type and the robustness of PID controlled loops depends on the tuning methodology and it is fact that the phase shaper proposed in this paper can be used in conjunction with PID controller thus making it a very useful tool. Further, the phase shaper proposed is of a low-order practically realizable system and the methodology allows the designer to cascade elements interactively starting from an initial point, established through a simple stability criterion. The extension of this methodology to analytically establish the adequacy of a single stage FO differ-integrator and its expression is left as a scope for future work.

## Acknowledgment

The work presented in this paper has been supported by the Board of Research in Nuclear sciences of the Department of Atomic Energy, India. Sanction no 2006/34/34-BRNS dated March 2007.